\documentclass[10pt,letterpaper,twocolumn]{article} 

\usepackage{ol2}
\usepackage{amsmath}

\begin{document}

\twocolumn[ 

\title{Off resonance laser frequency stabilization using the Faraday effect}

\author{Anna L. Marchant$^{*}$, Sylvi H\"{a}ndel, Timothy P. Wiles, Stephen A. Hopkins, Charles S. Adams and Simon L. Cornish$^{{\dagger}}$  }
\address{Department of Physics, Durham University, \\ Rochester Building, South Road, Durham DH1 3LE. United Kingdom\\
Corresponding authors: $^*$a.l.marchant@durham.ac.uk, $^{\dagger}$s.l.cornish@durham.ac.uk }

\begin{abstract}We present a simple technique for stabilization of a laser frequency off resonance using the Faraday effect in a heated vapor cell with an applied magnetic field. In particular we demonstrate stabilization of a 780~nm laser detuned up to 14~GHz from the $^{85}$Rb D$_2$ 5$^2$S$_{1/2}$ $F=2$ to 5$^2$P$_{3/2}$ $F'=3$ transition. Control of the temperature of the vapor cell and the magnitude of the applied magnetic field allows locking $\sim$6-14~GHz red and blue detuned from the atomic line. We obtain an rms fluctuation of 7~MHz over one hour without stabilization of the cell temperature or magnetic field.
\end{abstract}


] 


In the field of atomic and molecular physics it is commonplace to require a frequency stabilized or `locked' laser source. Many established techniques exist for locking close to an atomic resonance (within a Doppler linewidth) including: frequency-modulation spectroscopy \cite{FM}, polarization spectroscopy \cite{polspec}, dichroic atomic vapor laser locking \cite{Corwin:98} and Sagnac interferometry \cite{Robins:02} . In contrast, schemes for locking away from resonance, that is, greater than twice the Doppler linewidth away, can present more of an obstacle. Nevertheless, in many instances it is desirable to be detuned by more than 1~GHz from resonance. This is particularly important in experiments using two photon or Raman transitions where population of the intermediate state should be avoided. Examples include Raman cooling \cite{PhysRevLett.76.2658} and two photon excitation of Rydberg states \cite{PhysRevLett.100.113003}. It is also common to manipulate ultracold atoms in optical lattices formed using laser fields detuned many GHz from resonance. Applications of such lattices include degenerate Raman sideband cooling (DRSC) \cite{DRSC_kerman} and the study of quantum accelerator modes \cite{PhysRevLett.83.4447}.

A common approach for locking off resonance is to employ a second laser, slaved in some way to a reference laser which is locked on resonance. For example, an electro-optic modulator can be used to produce off resonant sidebands on the reference laser which are then used to seed the slave laser \cite{Bason2009}. Alternatively, a beat measurement can be used to stabilize the frequency difference between the slave and reference lasers \cite{Zielonkowski_99_beat}. In a similar fashion, an optical cavity can be used to bridge the frequency gap between the two lasers \cite{Hansch1980,Bohlouli_2006}. However, the experimental complexity of these approaches means that realising their full potential can be technically challenging, especially for larger detunings. Here we present a simple alternative based upon the Faraday effect in a heated vapor cell with an applied magnetic field which allows locking ~6-14 GHz (red and blue) detuned from an atomic line without the need for a second laser. 

The Faraday effect is a magneto-optical phenomenon. A magnetic field applied along the direction of light propagation causes the medium involved to respond differently to left and right circularly polarized light as the field shifts the $\sigma$ transitions.  The result of this circular birefringence is a rotation of linearly polarized light entering the medium due to the phase shift accumulated between its circular components. A key example of where this effect can be exploited is in optical isolators. A similar dispersive response to that exhibited by the isolator crystal can be seen in atomic media. 

In previous work the Faraday effect has been applied to produce a narrow-band optical filter or `Faraday filter' \cite{P.P.Sorokin_Farafilter} and, more recently, in a slow light medium to produce a gigahertz-bandwidth atomic probe \cite{Paul_nature}. In this letter we present a technique which uses the Faraday effect in a heated cell to lock off resonance from an atomic transition using that same transition. We consider the specific case of red detuning $>$10~GHz from the 780~nm $^{85}$Rb D$_2$ 5$^2$S$_{1/2}$ $F=2$ to 5$^2$P$_{3/2}$ $F'=3$ transition in order to perform degenerate Raman sideband cooling on ultracold atoms in the $F=2$ ground state.


\begin{figure}
	\centering
	\includegraphics{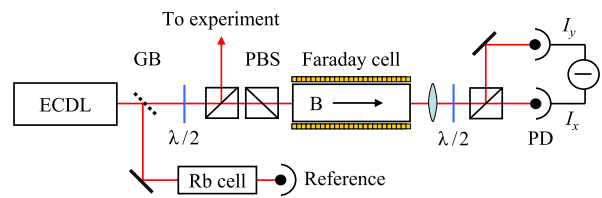}
			\caption{Schematic of the experimental apparatus. ECDL: external cavity diode laser, GB: glass blank, PBS: polarizing beam splitter, PD: differencing photodiode. }
		\label{fig:setup}
\end{figure}

\begin{figure}
	\centering
			\includegraphics{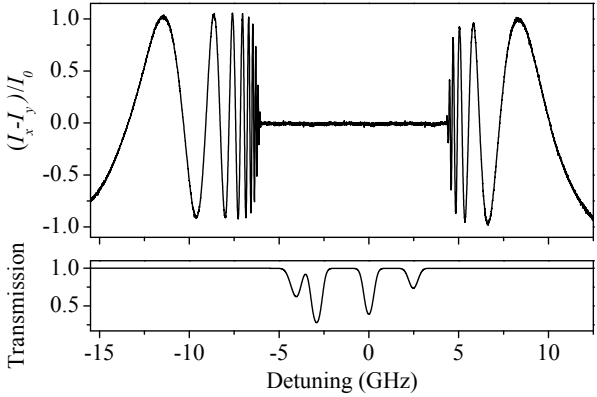}
			\caption{Normalized Faraday signal as a function of detuning from $^{85}$Rb $F=2$ to $F'=3$ obtained using a cell heated to 124$^{\circ}$C with $\sim$~270$~$G applied magnetic field. A room temperature Doppler broadened transmission spectrum is shown for reference.}
	\label{fig:bothsides}
\end{figure}

We use the setup shown in Fig. \ref{fig:setup} to produce the Faraday signal. Light is derived from a homebuilt 780~nm external cavity diode laser (based upon a Roithner Lasertechnik RLT 780-150 GS laser diode). The design incorporates current feed forward circuitry which allows the current to be modulated (with an amplitude of 20~mA) synchronously with the piezo. As a result of this modification the scan range of the laser is increased from 2.5~GHz to around 14~GHz. The laser output is split into two beams using a glass blank. The weak reflection from the glass blank is used to record a transmission spectrum using a room temperature vapor cell for reference. The transmitted beam is sent through a half-waveplate and polarizing beam splitter (PBS) cube to pick off light for the Raman sideband cooling, leaving a small proportion of the total output light for the Faraday beam. This light is sent through a second PBS to ensure well defined polarization (extinction of 500:1) and is attenuated to $\sim$~170~$\mu$W. Inside the cell the beam has a 1/e$^2$ radius of 1.30(2)~mm. The Faraday cell is a modification of the DAVLL cell used in \cite{DAVLL} consisting of two 43~mm long solenoids, each wound with 8 layers (53 turns per layer) of 0.8~mm polyurethane coated copper wire (rated at 150~$^{\circ}$C) surrounding a 7.5~cm long rubidium vapor cell. When supplied with 6~A the solenoids produce a magnetic field of $\sim$~270~G (the field varies smoothly between 240~G and 300~G along the length of the cell) and simultaneously heat the rubidium vapor cell enclosed to 110~$^{\circ}$C once thermal equilibrium is reached. The output from the Faraday cell is analysed by polarimetry. The light is split using a PBS and a half-waveplate set such that, in the absence of any optical rotation, an equal amount of light is incident on each of two photodiodes. (We note all polarization optics are positioned away from the cell to avoid thermal effects.) To produce the Faraday signal the difference in the two photocurrents (\textit{I$_x$ - I$_y$}) is converted to a voltage via a transimpedance amplifier ($R$ = 1.2~M$\Omega$). Fig. \ref{fig:bothsides} shows a typical signal normalized to the transmission in the absence of the applied magnetic field (obtained by switching off the cell current), $I_0$ = $I_{x_0}$ + $I_{y_0}$. Each zero crossing in the observed signal corresponds to a $\pi$ phase shift between the left and right circular components of the input light and is a potential locking point. A room temperature Doppler broadened transmission spectrum is shown to highlight that several lock points exist both red and blue detuned from resonance. Close to resonance there is no Faraday signal as the atomic vapor is optically thick.

\begin{figure}
	\centering
	\includegraphics{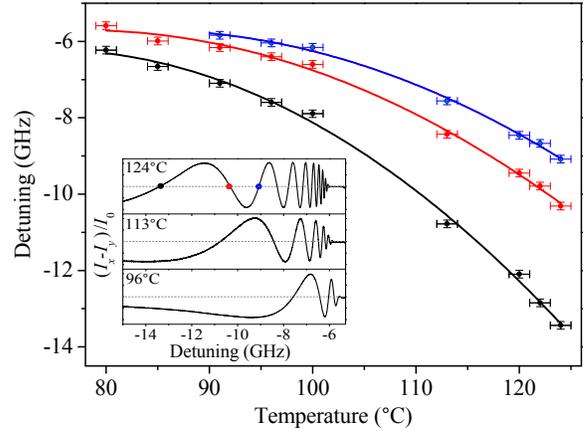}	
		\caption{Detuning from $^{85}$Rb $F=2$ to $F'=3$ of the last (black, bottom), second to last (red, center) and third to last (blue, top) zero crossings of the Faraday signal as a function of cell temperature for an applied magnetic field of $\sim$~270$~$G. At lower temperatures there is no visible third to last crossing. The fitted lines are a guide to the eye only. Inset: Example signals obtained with a magnetic field of $\sim$~270~G for three cell temperatures.}  
	\label{fig:temperature}
\end{figure}

The number and detuning of potential locking points exhibited by the Faraday signal are determined by the combination of the coil temperature and magnetic field. This, in principle, allows the detuning of the lock point to be controlled. To investigate the effect of temperature on the detuning of the lock points the cell was heated by adjusting the current supplied to the copper windings. The temperature was monitored using a K PTFE thermocouple probe inside the metal coil mounting around the cell. Before each measurement the current was set back to 6~A to generate the same magnetic field within the cell without allowing the temperature to change. As the temperature of the cell, and therefore the atomic vapor pressure, is increased, the refractive index of the medium increases leading to a greater circular birefringence and an enhanced magneto-optical effect at larger detunings. The detunings of the last, second to and third to last zero crossings with temperature are shown in Fig. \ref{fig:temperature}. This temperature dependence gives a good coarse adjustment of the lock point between $\sim$~6 and 14~GHz. Below $\sim$~70~$^{\circ}$C the rotation does not extend beyond the optically thick region and as such there are no suitable lock points. 

\begin{figure}
	\centering
	\includegraphics{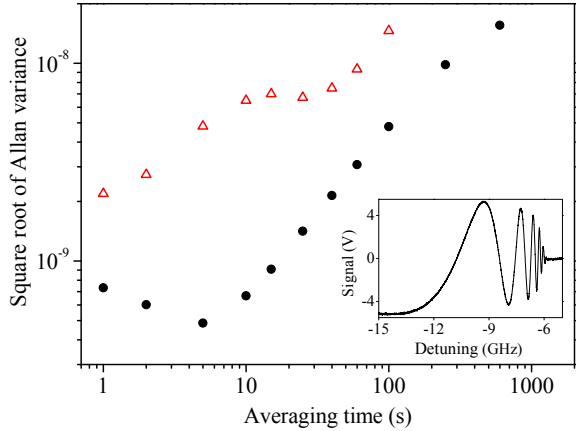}
		\caption{ Square root of the Allan variance of the beat frequency between two Faraday locked (circles) ECDLs. For comparison data for one free running laser and one locked with modulation transfer spectroscopy is also shown (triangles). Inset: Example raw photodiode signal obtained at 113$^{\circ}$C.}
	\label{fig:Allan}
\end{figure}

To lock to a zero crossing in the Faraday signal we use a homebuilt locking servo. The input to this is the non-normalized output of the differencing photodiode (see inset Fig. \ref{fig:Allan}). 
In order to investigate the stability of the Faraday lock a second copy of the setup shown in Fig. \ref{fig:setup} was constructed. Both lasers were locked using their respective Faraday signals and light taken from the two setups was combined and detected on a fast photodiode (2~GHz bandwidth). The frequency of the beat note between the lasers was recorded at 1~s intervals for $\sim$~80 minutes and the Allan variance \cite{Allan1966} calculated. Fig. \ref{fig:Allan} shows the square root of the Allan variance as a function of averaging time of the beat frequency between the two Faraday locked lasers. For comparison we also show the results for the case of one laser free running and the second locked with modulation transfer spectroscopy \cite{MT} to the $^{85}$Rb $F=3$ to $F'=4$ transition, stable to $\sim$~0.1~MHz. For the Faraday locked lasers the fractional frequency stability was 4.9~x~10$^{-10}$ for a 5~s averaging time and over the full monitor period the rms frequency deviation from the mean was $\sim$~7~MHz. For applications such as DRSC, where the resulting optical potential scales as 1/detuning, this frequency stability ($\approx$~0.1\% of the total detuning) is more than adequate.  

To understand the observed fluctuations the sensitivity of the signal to temperature and magnetic field was examined. From Fig. \ref{fig:temperature}, at 110~$^{\circ}$C, the temperature dependence of the last zero crossing is $\sim$~-0.2~GHz/$^{\circ}$C. The effect of magnetic field gives a shift of -~0.49(2)~MHz/mA, which, assuming a field of 270~G translates into a field sensitivity of -~10.9(4)~MHz/G. Considering these sensitivities we attribute the observed fluctuations to small changes in the cell temperature caused by the current supply. We note that the supply to the coil was operated in constant current mode but the current was not actively stabilized. In addition, there was no active stabilization of the ambient temperature and magnetic field. However, with stabilization of temperature ($\sim$~1~mK) and current ($\sim$~10~$\mu$A), frequency stability on the order of 200~kHz should be possible.

In summary, we have demonstrated a simple laser locking technique which allows off resonance ($\sim$~6-14~GHz red and blue detuned) stabilization to an absolute frequency using the Faraday effect. A compact heated cell is used, not only to achieve increased rubidium vapor pressure but also to produce a magnetic field parallel to the direction of light propagation. Implementation of the Faraday locking technique extends the useful locking range of an atomic reference considerably. The usual accessible sub Doppler transition range of a few hundred MHz is increased to, in our case, around 25~GHz. We propose the technique for use in the generation of off resonant optical lattices, such as those needed in DRSC schemes. Potentially the detuning achieved could be extended by careful cell design. From extrapolation of Fig. \ref{fig:temperature}, we expect detunings of 25~GHz and 50~GHz would require 155~$^{\circ}$C and 200~$^{\circ}$C operating temperatures respectively. Cell designs capable of achieving such temperatures are currently under construction.


We thank P. Siddons for useful discussions. We acknowledge support from the UK EPSRC and the ESF within the EUROCORES Programme EuroQUASAR. SLC acknowledges the support of the Royal Society.



\end{document}